\documentclass[a4paper,twocolumn,english,prl,showpacs,preprintnumbers,amsmath,amssymb]{revtex4}
\usepackage[T1]{fontenc}
\usepackage[latin9]{inputenc}
\usepackage{graphicx}
\usepackage{amssymb}

\usepackage{babel}

\begin{document}

\title{Universality and Anomalous Mean-Field Breakdown of Symmetry-Breaking
Transitions in A Coupled Two-Component Condensate}

\author{Chaohong Lee}

\altaffiliation{Electronic addresses: chl124@rsphysse.anu.edu.au; chleecn@gmail.com}

\affiliation{Nonlinear Physics Centre and ARC Centre of Excellence for Quantum-Atom
Optics, Research School of Physical Sciences and Engineering, Australian
National University, Canberra ACT 0200, Australia}

\date{\today}
\begin{abstract}
We study both mean-field and full quantum dynamics of symmetry-breaking
transitions (SBTs) in a coupled two-component Bose-Einstein condensate.
By controlling s-wave scattering lengths and coupling strength, it
is possible to stimulate SBTs between normal and spontaneously polarized
ground states. In static transitions, the probability maxima of full
quantum ground states correspond to the mean-field ground states.
In dynamical transitions, due to the vanishing of excitation gaps,
the mean-field dynamics shows universal scalings obeying Kibble-Zurek
mechanism. Both mean-field and full quantum defect modes appear as
damped oscillations, but they appear at different critical points
and undergo different oscillation regimes. The anomalous breakdown
of mean-field dynamics induced by SBTs depends on the approaching
direction.
\end{abstract}


\maketitle
In a many-body quantum system, spontaneous symmetry breaking (SSB)
occurs if its mean-field (MF) states do not possess symmetry of its
original many-body Hamiltonian. Since atomic Bose-Einstein condensates
(BECs) have long collision times, they are excellent candidates for
testing intrinsic mechanisms of the SSB~\cite{Ueda-AIP_Conf_Proc}.
In particular, the simultaneous realization of superfluidity and magnetism
(or spin polarization) in spinor BECs is associated with the SSB related
to both external and internal degrees of freedom. For spin-1 BECs,
the experimental observation of SSB~\cite{Nature-SSB-spinor-BEC}
has triggered several theoretical studies~\cite{Ueda-SSB-spinor-BEC,Damski-SSB-spinor-BEC}.
For spin-$\frac{1}{2}$ (two-component) BECs, the phase separation
has been observed experimentally~\cite{Two-component-BEC}, and their
spatial SSB~\cite{Esry-spatial-SSB} and spontaneous spin polarization~\cite{Lee-SSP}
have been predicted theoretically.

It is well known that, in slow processes, the gapped excitations over
zero-temperature ground states obey Landau-Zener mechanisms. However,
gapless excitations appear in almost all systems with SSB. Because
of the gapless excitations, the adiabaticity breaks down~\cite{Polkovnikov-BAL}
and the generated defect modes~\cite{Ueda-SSB-spinor-BEC,Damski-SSB-spinor-BEC,DQPT-KZ-mechanism}
follow Kibble-Zurek (KZ) mechanisms~\cite{KZ-mechanism}. The current
studies of dynamical mechanisms for SBTs focus on lattice systems~\cite{Polkovnikov-BAL,DQPT-KZ-mechanism,PRL-Josephson-junctions-array,PRL-SF-MI-BHM},
spin-1 BECs~\cite{Ueda-SSB-spinor-BEC,Damski-SSB-spinor-BEC} and
other many-body systems by employing either MF~\cite{Damski-SSB-spinor-BEC,PRL-Josephson-junctions-array}
or full quantum (FQ)~\cite{Ueda-SSB-spinor-BEC,Polkovnikov-BAL,DQPT-KZ-mechanism,PRL-SF-MI-BHM}
theories. However, (i) few works compare the MF and FQ dynamical mechanisms
near a critical point to explore regimes of correspondence, and (ii)
the dynamical mechanisms for SBTs in two-component BECs are still
not clearly understood.

In this Letter, we analyze both MF and FQ dynamical mechanisms for
SBTs in a coupled two-component BEC. For simplicity, we only consider
the SSB related to internal degrees of freedom, i.e., two hyperfine
spin states. In static transitions, the MF ground states correspond
to the probability maxima of the FQ ground states. If the coupling
strength increases from zero to a sufficiently large quantity, the
MF ground states transfer from being spontaneously polarized (self-trapped)
to non-polarized (normal). Correspondingly, the two lowest FQ eigenstates
transfer from being quasi-degenerated to non-degenerated. In dynamical
transitions, due to the disappearance of gaped Bogoliubov excitations,
the MF dynamics obeys an universal KZ mechanism and the defect modes
are induced by dynamical instability. Due to the non-identity of the
FQ and MF critical points, the MF breakdown induced by SBTs depends
on the approaching direction. The dynamical mechanism of SBTs connects
with the quantum adiabaticity, which provides various applications
in atomic physics, condensed matter physics and non-equilibrium dynamics,
and particularly in adiabatic quantum computation~\cite{Farhi-Science}.

We consider a gaseous BEC of bosonic atoms which populate two hyperfine
levels coupled by external fields. Assuming that the coupling fields
are spatial uniform and the atom-atom interactions do not change the
internal states, the system obeys a second quantized Hamiltonian~\cite{Nature-SMA},
\begin{eqnarray*}
H(t) & = & H_{0}+H_{int}+H_{c}(t),\\
H_{0} & = & \sum_{j=\uparrow,\downarrow}\int\hat{\Psi}_{j}^{+}(\vec{r})\left(-\frac{\hbar^{2}\nabla^{2}}{2m}+V+\epsilon_{j}\right)\hat{\Psi}_{j}(\vec{r})d^{3}\vec{r},\\
H_{int} & = & \frac{1}{2}\sum_{j,k=\uparrow,\downarrow}U_{jk}\int\hat{\Psi}_{j}^{+}(\vec{r})\hat{\Psi}_{k}^{+}(\vec{r})\hat{\Psi}_{k}(\vec{r})\hat{\Psi}_{j}(\vec{r})d^{3}\vec{r},\\
H_{c}(t) & = & -\frac{\hbar\Omega(t)}{2}\int\left(\hat{\Psi}_{\uparrow}^{+}(\vec{r})\hat{\Psi}_{\downarrow}(\vec{r})+\hat{\Psi}_{\downarrow}^{+}(\vec{r})\hat{\Psi}_{\uparrow}(\vec{r})\right)d^{3}\vec{r}.\end{eqnarray*}
 Here, $m$ is the single-atom mass, $V=\frac{1}{2}m\omega^{2}r^{2}$
is the trapping potential, $\Omega(t)\geq0$ is the coupling strength,
$\epsilon_{j}$ are the hyperfine energies, $\hat{\Psi}_{j}(\vec{r})$
are the field operators, and $U_{jk}=U_{kj}=4\pi\hbar^{2}a_{jk}/m$
are interaction strengths of atoms in states $|j\rangle$ and $|k\rangle$
parameterized by s-wave scattering lengths $a_{jk}$. To analyze dynamics
only associated with internal degrees of freedom, we apply a single-mode
approximation~\cite{Nature-SMA}: $\hat{\Psi}_{j}(\vec{r})=\hat{b}_{j}\phi(\vec{r})$,
that is, assume that all atoms occupy the same single-particle external
state $\phi(\vec{r})$. Thus the Hamiltonian is simplified as follows:
\begin{eqnarray}
H & = & -\frac{\hbar\Omega(t)}{2}\left(\hat{b}_{\uparrow}^{+}\hat{b}_{\downarrow}+\hat{b}_{\downarrow}^{+}\hat{b}_{\uparrow}\right)+G_{\uparrow\downarrow}\hat{b}_{\uparrow}^{+}\hat{b}_{\downarrow}^{+}\hat{b}_{\downarrow}\hat{b}_{\uparrow}\nonumber \\
 & ~ & +\sum_{j=\uparrow,\downarrow}^{}\left(E_{0j}\hat{b}_{j}^{+}\hat{b}_{j}+\frac{1}{2}G_{jj}\hat{b}_{j}^{+}\hat{b}_{j}^{+}\hat{b}_{j}\hat{b}_{j}\right),\end{eqnarray}
 where $E_{0j}=\int\phi^{*}(\vec{r})\left(-\frac{\hbar^{2}\nabla^{2}}{2m}+V(\vec{r})+\epsilon_{j}\right)\phi(\vec{r})d^{3}\vec{r}$
and $G_{jk}=U_{jk}\int\left|\phi(\vec{r})\right|^{4}d^{3}\vec{r}$.
This simplification can successfully describe the system of uniform
polarization. The Hamiltonian is equivalent to a Bose-Josephson junction~\cite{BJJ-Sols,BJJ-Lee,BJJ-review},
which can be realized by a BEC in a double-well trap~\cite{DW-BEC}.
In contrast to the double-well systems, in which negative charging
energies $E_{C}\propto a_{s}$ may cause spatial collapse, the two-component
systems support negative $E_{C}\propto(a_{\uparrow\uparrow}+a_{\downarrow\downarrow}-2a_{\uparrow\downarrow})$
with all positive $a_{jk}$.

In MF theory, by introducing variational trial states $|\Phi\rangle$
as coherent states~\cite{JPA-Penna}, $\psi_{j}^{*}=\langle\Phi|\hat{b}_{j}^{+}|\Phi\rangle$
and $\psi_{j}=\langle\Phi|\hat{b}_{j}|\Phi\rangle$ obey the classical
Hamiltonian: \begin{eqnarray}
H_{MF} & = & -\frac{\hbar\Omega(t)}{2}\left(\psi_{\uparrow}^{*}\psi_{\downarrow}+\psi_{\downarrow}^{*}\psi_{\uparrow}\right)+G_{\uparrow\downarrow}\left|\psi_{\uparrow}\right|^{2}\left|\psi_{\downarrow}\right|^{2}\nonumber \\
 & ~ & +\sum_{j=\uparrow,\downarrow}\left(E_{0j}\left|\psi_{j}\right|^{2}+\frac{1}{2}G_{jj}\left|\psi_{j}\right|^{4}\right).\end{eqnarray}
 Rewriting $\psi_{j}=\sqrt{N_{j}}\exp(i\theta_{j})$ in terms of the
particle numbers $N_{j}$ and phases $\theta_{j}$, we denote $l=(N_{\uparrow}-N_{\downarrow})/2$,
$L=(N_{\downarrow}+N_{\uparrow})/2$, $\Delta=E_{0\uparrow}-E_{0\downarrow}+L(G_{\uparrow\uparrow}-G_{\downarrow\downarrow})$
and $E_{C}=G_{\uparrow\uparrow}+G_{\downarrow\downarrow}-2G_{\uparrow\downarrow}$.
It has been predicted that spontaneous spin polarization, a type of
SSB related to self-trapping and bi-stability~\cite{BJJ-Sols,MQST},
can appear when $E_{C}<0$~\cite{Lee-SSP}. For symmetric systems
($\Delta=0$) of negative $E_{C}$, the SBT occurs if $|\hbar\Omega/E_{C}|$
varies from $|\hbar\Omega/E_{C}|>L$ to $|\hbar\Omega/E_{C}|<L$.
Correspondingly, the MF ground state changes from $(\psi_{\downarrow},\psi_{\uparrow})=(\sqrt{L},\sqrt{L})\exp\left(i\theta_{\downarrow}\right)$
to $(\sqrt{L-l_{s}},\sqrt{L+l_{s}})\exp\left(i\theta_{\downarrow}\right)$,
where $l_{s}=\pm\sqrt{L^{2}-\hbar^{2}\Omega^{2}/E_{C}^{2}}$. Below,
we will focus on the SBTs in symmetric systems of $E_{C}<0$.

First, let us analyze the MF-FQ correspondence for static SBTs. It
is obvious that the FQ ground state appears as an SU(2) coherent state
if $E_{C}=0$, and the ground and first excited states become degenerate
if $\Omega=0$ and $E_{C}<0$. For other arbitrary parameters, our
numerical results show that the FQ ground states are always symmetric
with respect to $l=0$ and the probability distributions change from
single-hump shapes to double-hump ones when $\hbar\Omega/E_{C}$ passes
the critical point $\hbar\Omega/E_{C}=-L$. However, the first excited
states are always anti-symmetric with respect to $l=0$ and the probability
distributions retain double-hump shapes.

The probability distributions and degeneracy properties of the low-energy
FQ states are reminiscent of a single quantum particle confined within
a potential that varies from single-well to double-well configuration.
By comparing the FQ and MF ground states, it is easy to find that
the MF ground states correspond to the probability maxima of the FQ
ones. Because the quasi-degeneracy between the two lowest FQ states
requires almost identical probability distributions for these two
states, the MF bifurcation point is not identical to the degeneracy
point for the two lowest FQ eigenstates. In Fig.~1, we demnstrate
the MF-FQ correspondence for the static SBT in a symmetric system
of total particle number $N=N_{\downarrow}+N_{\uparrow}=100$ and
$E_{C}<0$. The supercritical pitchfork bifurcation of the MF stationary
states signals the static SBT. Similar bifurcations have also been
found in rotating BECs~\cite{PRL-Sinha}. Due to the appearance or
disappearance of unstable or stable stationary states, the bifurcation
could cause dynamical instability.

\begin{figure}[ht]
\includegraphics[width=1\columnwidth]{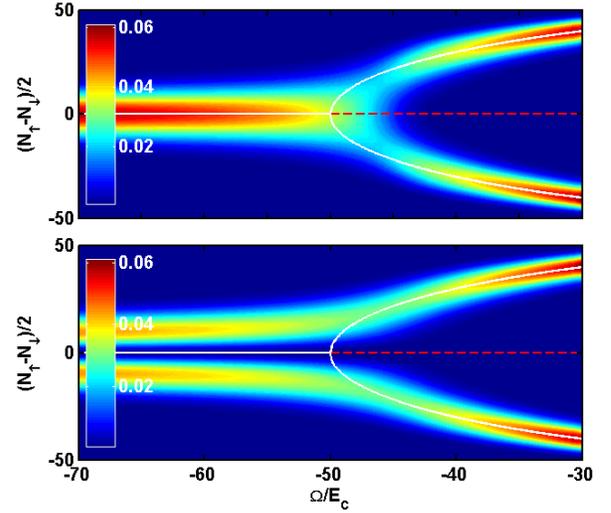}

\caption{The MF-FQ correspondence for static SBTs in a coupled two-component
condensate of $N=100$ and negative $E_{C}$. The FQ states are defined
as $|\Psi\rangle=\sum_{l=-L}^{L}C(l)|l\rangle$ and the probability
distributions $|C(l)|^{2}$ versus $\Omega/E_{C}$ for the ground
(first-excited) states are shown in the above (bottom) panel. The
solid and dashed lines are stable and unstable MF stationary states,
respectively.}

\end{figure}

Now, let us consider the dynamical SBTs. In a dynamical process, there
exist two characteristic time scales~\cite{Damski-SSB-spinor-BEC}.
One time scale is the \textit{reaction time}, $\tau_{r}(t)=\hbar/\Delta_{g}(t)$,
which characterizes how fast the system follows eigenstates of its
instantaneous Hamiltonian. Here, $\Delta_{g}(t)$ is the instantaneous
excitation gap over the ground state. The other timescale is the \textit{transition
time}, $\tau_{t}(t)=\Delta_{g}(t)/\left|d\Delta_{g}(t)/dt\right|$,
which tells us how fast the system is driven. If and only if $\tau_{r}<\tau_{t}$,
the system undergoes adiabatic evolution.

To obtain the excitation modes over MF ground states, we perform a
Bogoliubov analysis. Apart from a trivial gapless mode, there exists
a gapped mode: \begin{equation}
\Delta_{g}=\left\{ \begin{array}{ll}
\hbar\sqrt{\Omega\left(\Omega-\Omega_{c}\right)},\textrm{~for~}\Omega\geq\Omega_{c},\\
\hbar\sqrt{\Omega_{c}^{2}-\Omega^{2}},\textrm{~for~}\Omega\leq\Omega_{c},\end{array}\right.\end{equation}
 where $\Omega_{c}=\left|E_{C}L/\hbar\right|$. Obviously, the gap
$\Delta_{g}$ gradually vanishes when $\Omega$ approaches the critical
point $\Omega_{c}$.

To analyse the dynamical mechanism, we suppose $\Omega(t)=\Omega_{c}\left(1\pm t/\tau_{q}\right)=\Omega_{c}\pm\beta t$,
where $\tau_{q}$ denotes the \textit{quenching time}. The relative
coupling, $\varepsilon=\left|\left(\Omega(t)-\Omega_{c}\right)/\Omega_{c}\right|=\left|t\right|/\tau_{q}$,
corresponds to the relative temperature in KZ theory~\cite{KZ-mechanism}.
Due to the gap $\Delta_{g}\rightarrow0$ when $\left|\Omega-\Omega_{c}\right|\rightarrow0$,
the adiabaticity breaks down when $\Omega\rightarrow\Omega_{c}$ and
then revives after $\Omega$ passes $\Omega_{c}$ {[}see panel (a)
of Fig. 2]. The time of $\tau_{r}\left(\hat{t}\right)=\tau_{t}\left(\hat{t}\right)$,
where adiabatic-diabatic transition occurs, corresponds to the \emph{freeze-out
time} in KZ theory~\cite{KZ-mechanism}. In the single-stable region,
$\Omega>\Omega_{c}$, introducing $\hat{\Omega}(t)=\Omega(t)-\Omega_{c}$,
we have $\tau_{r}=\frac{1}{\sqrt{\hat{\Omega}\left(\hat{\Omega}+\Omega_{c}\right)}}$
and $\tau_{t}=\frac{2\hat{\Omega}\left(\hat{\Omega}+\Omega_{c}\right)}{\Omega_{c}\left(2\hat{\Omega}+\Omega_{c}\right)}\tau_{q}$.
By solving $\tau_{r}\left(\hat{t}\right)=\tau_{t}\left(\hat{t}\right)$,
we obtain the relation \begin{equation}
\tau_{q}=\frac{\left(2\hat{\Omega}+\Omega_{c}\right)\Omega_{c}}{2\left[\hat{\Omega}\left(\hat{\Omega}+\Omega_{c}\right)\right]^{3/2}}=\frac{2\varepsilon+1}{2\left[\varepsilon\left(\varepsilon+1\right)\right]^{3/2}}\tau_{0},\end{equation}
 where $\tau_{0}=1/\Omega_{c}$. For slow transitions, $\tau_{q}\gg1$,
we find \begin{equation}
\left|\hat{t}\right|\approx2^{-2/3}\tau_{0}^{2/3}\tau_{q}^{1/3},\quad\varepsilon\approx2^{-2/3}\tau_{0}^{-2/3}\tau_{q}^{-2/3}.\end{equation}
 In the bi-stable region, $\Omega<\Omega_{c}$, introducing $\hat{\Omega}(t)=\left.\Omega_{c}-\Omega(t)\right.$,
we have $\tau_{r}=\frac{1}{\sqrt{\hat{\Omega}\left(2\Omega_{c}-\hat{\Omega}\right)}}$
and $\tau_{t}=\frac{\hat{\Omega}\left(2\Omega_{c}-\hat{\Omega}\right)}{\Omega_{c}\left(\Omega_{c}-\hat{\Omega}\right)}\tau_{q}$.
Similarly, at the \emph{freeze-out time} $\hat{t}$, we have \begin{equation}
\tau_{q}=\frac{\left(\Omega_{c}-\hat{\Omega}\right)\Omega_{c}}{\left[\hat{\Omega}\left(2\Omega_{c}-\hat{\Omega}\right)\right]^{3/2}}=\frac{1-\varepsilon}{\left[\varepsilon\left(2-\varepsilon\right)\right]^{3/2}}\tau_{0}.\end{equation}
 For slow transitions, it is easy to find$\left|\hat{t}\right|\approx\frac{1}{2}\tau_{0}^{2/3}\tau_{q}^{1/3}$
and $\varepsilon\approx\frac{1}{2}\tau_{0}^{-2/3}\tau_{q}^{-2/3}$. 

The universal scalings $\left|\hat{t}\right|\sim\tau_{0}^{2/3}\tau_{q}^{1/3}$
and $\varepsilon\sim\tau_{0}^{-2/3}\tau_{q}^{-2/3}$ recover the KZ
mechanisms: $\left|\hat{t}\right|\sim\tau_{0}^{1/(1+z\nu)}\tau_{q}^{z\upsilon/(1+z\nu)}$
and $\varepsilon\sim\tau_{0}^{-1/(1+z\nu)}\tau_{q}^{-1/(1+z\nu)}$
with $z=1$ and $\nu=1/2$, for continuous quantum phase transitions~\cite{Damski-SSB-spinor-BEC,KZ-mechanism}.

Due to the adiabatic condition becoming invalid near the critical
point, the defect modes could be stimulated. In Fig. 2, we show the
dynamics of SBTs in the MF system. In our simulation, the initial
state is chosen as a symmetry-broken ground state in the bi-stable
region, and $\Omega$ is ramped up from below to above $\Omega_{c}$.
For different ramping rates, we calculate the longitudinal polarization
and the fidelity to instantaneous ground states. The results show
that the MF defect modes appear as damped oscillations with $\beta$-dependent
amplitudes {[}see panels (b) and (c) of Fig. 2]. The defect generation
via dynamical insatiability is similar to the vortex nucleation in
rotating BECs~\cite{PRL-Sinha}. Interestingly, the defect modes
become significant after the revival of adiabaticity when the system
has passed the critical point, but not after the breakdown of adiabaticity
when the system approaches the critical point. This indicates that
the system has no sufficient time to complete a whole intrinsic oscillation
in the interval between the breakdown and revival of adiabaticity.
Because of this, the system tries to keep its present state when the
adiabatic condition is invalid.

\begin{figure}[ht]
\includegraphics[width=1\columnwidth]{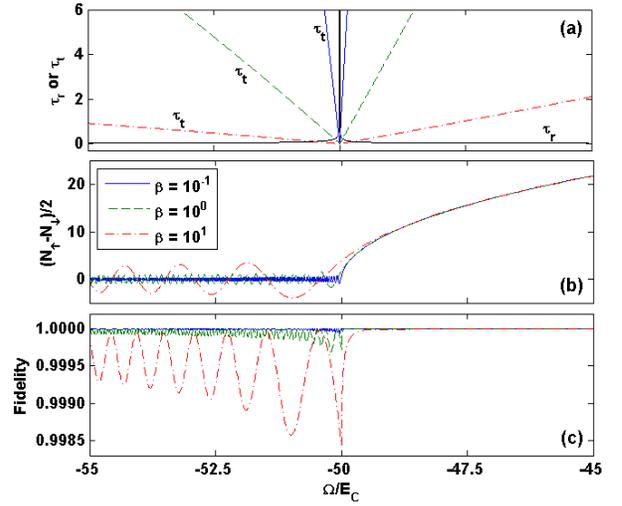}

\caption{The MF dynamics of SBTs in Hamiltonian (2). (a) The $\tau_{t}$ and
$\tau_{r}$, (b) the longitudinal polarization $(N_{\uparrow}-N_{\downarrow})/2$,
and (c) the fidelity to instantaneous ground states versus $\Omega/E_{C}$
for different values of $\beta$. Parameters are chosen as $N=100$,
$G_{\downarrow\downarrow}=G_{\uparrow\uparrow}=1.0$, $G_{\uparrow\downarrow}=2.0$,
and $E_{0\downarrow}=E_{0\uparrow}$. So that, $E_{C}=-2.0$ and $\Delta=0$.}

\end{figure}

To explore the correspondence between MF and FQ dynamics of SBTs,
we compare the dynamics of Hamiltonians (1) and (2). In Fig. 3, we
show the FQ dynamics corresponding to Fig. 2. In the FQ system, similar
defect modes appear as damped oscillations. However, their critical
points and oscillation regimes are different from the MF ones. First,
the FQ defect modes appear after the critical point between quasi-degeneracy
and non-degeneracy, which is not identical to the MF critical point
between bi-stability and single-stability {[}see panel (b) of Fig.
3]. Additionally, the FQ defect modes have $\beta$-independent amplitudes
{[}see panel (c) of Fig. 3]. Furthermore, starting from a ground state
in the single-stable region, the MF defect modes will always appear
if the system passes its critical point. However, if $|\beta|$ is
sufficiently small, there is no FQ defect mode. That is, the FQ system
always remains in its instantaneous ground state. This means that,
the MF dynamics well coincides with the FQ one before the system reaches
the first critical point and then it breaks down. The MF breakdown
occurs at its FQ critical point if the system goes from bi-stability
to single-stability. In contrat, the MF breakdown occurs at its MF
critical point if the system goes from single-stability to bi-stability.
This anomalous MF breakdown dependent on approaching direction differs
from the conventional MF breakdown related to interaction strength
and total particle number.

\begin{figure}[ht]
\includegraphics[width=1\columnwidth]{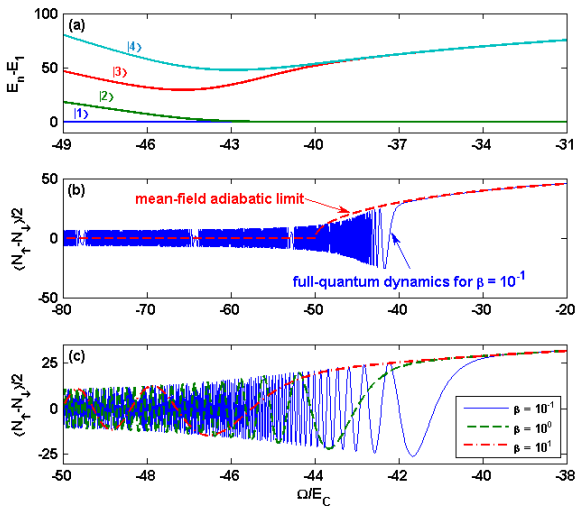}

\caption{The energy spectrum and FQ dynamics of SBTs shown in Fig. 2. (a) Energy
spectrum of four lowest states of Hamiltonian (1), (b) and (c) quantum
expectations of the longitudinal polarization $\langle N_{\uparrow}-N_{\downarrow}\rangle/2$.}

\end{figure}

To observe the dynamical SBTs, one should prepare a two-component
BEC~\cite{Two-component-BEC}. The negative $E_{C}$ with all positive
s-wave scattering lengths can be obtained via Feshbach resonances.
To avoid the phase separation induced by strong inter-component repulsion,
one has to trap the BEC within a sufficiently strong potential. The
inter-component coupling can be realized by Raman and/or radio-frequency
fields, and then the coupling strength is controlled by the field
intensity. By ramping the field intensity up or down, one can push
the system pass its critical points. To detect the longitudinal polarization,
one should count the atoms in each component via state-dependent fluorescence
or spatial imaging.

In conclusion, we have studied both MF and FQ dynamics of SBTs in
a coupled two-component BEC. We analytically obtain universal KZ scalings
for MF dynamics and numerically explore the correspondence between
MF and FQ dynamics. The MF dynamics well coincides with the FQ one
before reaching the first critical point and then it breaks down.
Because the FQ critical point is not identical to the MF one, the
MF breakdown induced by SBTs depends on the approaching direction.
In a transition from the normal to polarized regions, the MF breakdown
occurs at the MF critical point. However, in a transition from the
polarized to normal regions, the MF breakdown occurs at the FQ critical
point. This anomalous MF breakdown broadens our conventional understanding
for MF breakdown related to interaction strength and total particle
number.

The author acknowledges valuable discussions with Elena A. Ostrovskaya
and Yuri S. Kivshar. This work is supported by the Australian Research
Council (ARC).

\end{document}